\documentclass[%
 aip,
 sd,%
 amsmath,amssymb,
 reprint,%
]{revtex4-1}
\usepackage{graphicx}
\usepackage{dcolumn}
\usepackage{bm}
\usepackage{upgreek}

\begin{document}

\preprint{AIP/123-QED}

\title[]{Optical nanofiber-based cavity induced by periodic air-nanohole arrays}     
\author{Wenfang Li}
 \email{wenfang.li@oist.jp}
\author{Jinjin Du}
\author{Viet Giang Truong}
\author{S\'ile Nic Chormaic}
\affiliation{%
Light-Matter Interactions Unit, Okinawa Institute of Science and Technology Graduate University, Onna,Okinawa 904-0495, Japan
}%

\date{\today}

\begin{abstract}
We experimentally realized an optical nanofiber-based cavity by combining a 1-D photonic crystal and Bragg grating structures. The cavity morphology comprises a periodic, triplex air-cube introduced at the waist of the nanofiber. The cavity has been theoretically characterized using FDTD simulations to obtain the reflection and transmission spectra. We have also experimentally measured the transmission spectra and a Q-factor of $\sim$ $784 \pm 87$ for a very short periodic structure has been observed. The structure provides strong confinement of the cavity field and its potential for optical network integration makes it an ideal candidate for use in nanophotonic and quantum information systems.
\end{abstract}

\maketitle

 Strong, efficient light-matter interactions are essential for quantum computing and quantum communication systems. To reduce complexity, and to compensate for the general unsatisfactory scalability of light-matter systems in free space, optical nanofibers (ONF) are proving to be a very useful tool for hybrid quantum devices wherein quantum emitters are coupled to the fiber-guided modes via the evanescent field extending from the nanofiber surface. In recent years, significant research effort has been invested in order to interface various quantum emitters, including neutral atoms\cite{1,2,3,4,5,6,7}, semiconductor quantum dots\cite{8,9}, and nitrogen-vacancy centers in diamond\cite{10}, to the light fields of ONFs. With tight, transverse-mode confinement and the relatively high intensity of the evanescent field, efficient light-matter coupling\cite{11} and low optical power nonlinearities\cite{12,13,14} have been experimentally demonstrated.

It is well-known that the strength of light-matter interactions can be further augmented by confining photons in an optical cavity\cite{15}, and the same is true if the cavity is fabricated from micro- or nanofibers. Various nanofiber-based resonator structures have been designed, such as the fiber Bragg grating (FBG) cavity\cite{16,17} or the photonic crystal (PhC) cavity\cite{18}.  Wuttke et al. reported the construction of a  Fabry-P\'erot type optical microresonator on a tapered ONF by linking two normal fiber Bragg grating mirrors\cite{19}. Similarly, Kato and Aoki experimentally observed strong coupling between single atoms and a moderate-finesse cavity in an all-fiber cavity system\cite{20}. A nanofiber Bragg grating cavity can also be made directly using a focused ion beam (FIB) technique, which can modify or mill the fiber surface with nanometer precision. This technique has been  used to fabricate various structures on micro/nanofibers\cite{21,22,23}. Furthermore, introducing photonic crystal structures into optical nanofibers also allows cavity formation, such as was done via the femtosecond laser ablation method for producing a nanocrater array\cite{24}. Keloth et al. used this technique  to create a centimeter-long cavity on a nanofiber and they proposed that the system could be used for strong-coupling  cavity quantum electrodynamics (cQED) \cite{25}. A composite photonic crystal cavity based on a nanofiber has also been formed by means of an external grating\cite{26}, and has been used to observe a significant enhancement of the spontaneous emission rate of single quantum dots into nanofiber guided modes\cite{8,9,27}. In addition, a fiber ring cavity containing a nanofiber section that strongly couples atoms to the cavity mode has been studied recently\cite{28,29}.

In this letter, we investigate a nanofiber-based cavity that incorporates both FBG and PhC structures in order to exploit the advantages of both simultaneously. Mirrors in this cavity exhibit higher reflectivity than other reported nanofiber cavities that are based solely on Bragg gratings or on 1-D photonic crystal structures. Accordingly, we can obtain a higher quality factor for the FBG-PhC cavities under similar conditions. This paves the way for studying optical nonlinear effects and cavity QED using nanofiber-based cavities.

Figure 1 shows a schematic diagram of the nanofiber cavity, which is sandwiched between two cavity mirrors, fabricated by etching periodic structures on the nanofiber. For ONFs, we explore three cavity mirror morphologies: Type I: 1-D photonic crystal, Type II: Bragg grating, and Type III: a combination of both. Figure 1 also illustrates (grey, ``cavity region") the fabricated section with a cavity length equal to the distance between the two mirror inner edges. Fig. 1 (b) shows a Type III structure. The parameters to consider include the  etch width, $a$, etch depth, $b$, period, $c$, and etch number, $N$,  for a nanofiber of diameter $D$. The refractive index for the ${\rm{Si}}{{\rm{O}}_{\rm{2}}}$ air-clad fiber is 1.45. The wavelength band of the cavity is determined by the period, $c$ and exhibits a red-shift as $c$ increases. The etch width and depth influence the spatial modulation depth of the fiber's effective refractive index. The depth of the wavelength band increases proportionally to the etch area $(a \times b)$. The parameters are, however, restricted by experimental limitations. 

\begin{figure}[htb]
\centerline{
\includegraphics[width=8cm]{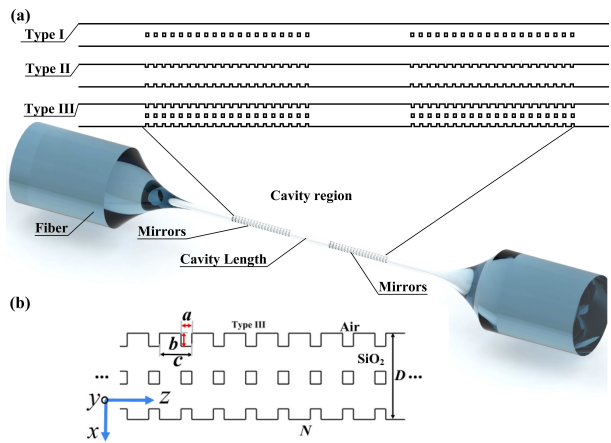}}
 \caption{(a) Schematics of three different ONF cavity structures (Types I, II, and III). (b) Design of a Type III cavity. Indicated are the cavity parameters, etch width $a$, etch depth $b$, period $c$, and etch numbers $N$ for a nanofiber of diameter $D$.}
\end{figure}

We have characterized the transmission and reflection of the three ONF cavity types using a finite-difference, time-domain method (FDTD, Photon Design Inc.), given by figure 2. A resonance wavelength close to the rubidium ${D_2}$ line was chosen (near 780 nm), in accordance with our experimental conditions. The parameters were defined as follows: $a = b = 100$ nm; $c = 310 $ nm, and $N$ was varied from $5$ to $30$ with a step size of $5$.

\begin{figure}[htb]
\centerline{
\includegraphics[width=8.5cm]{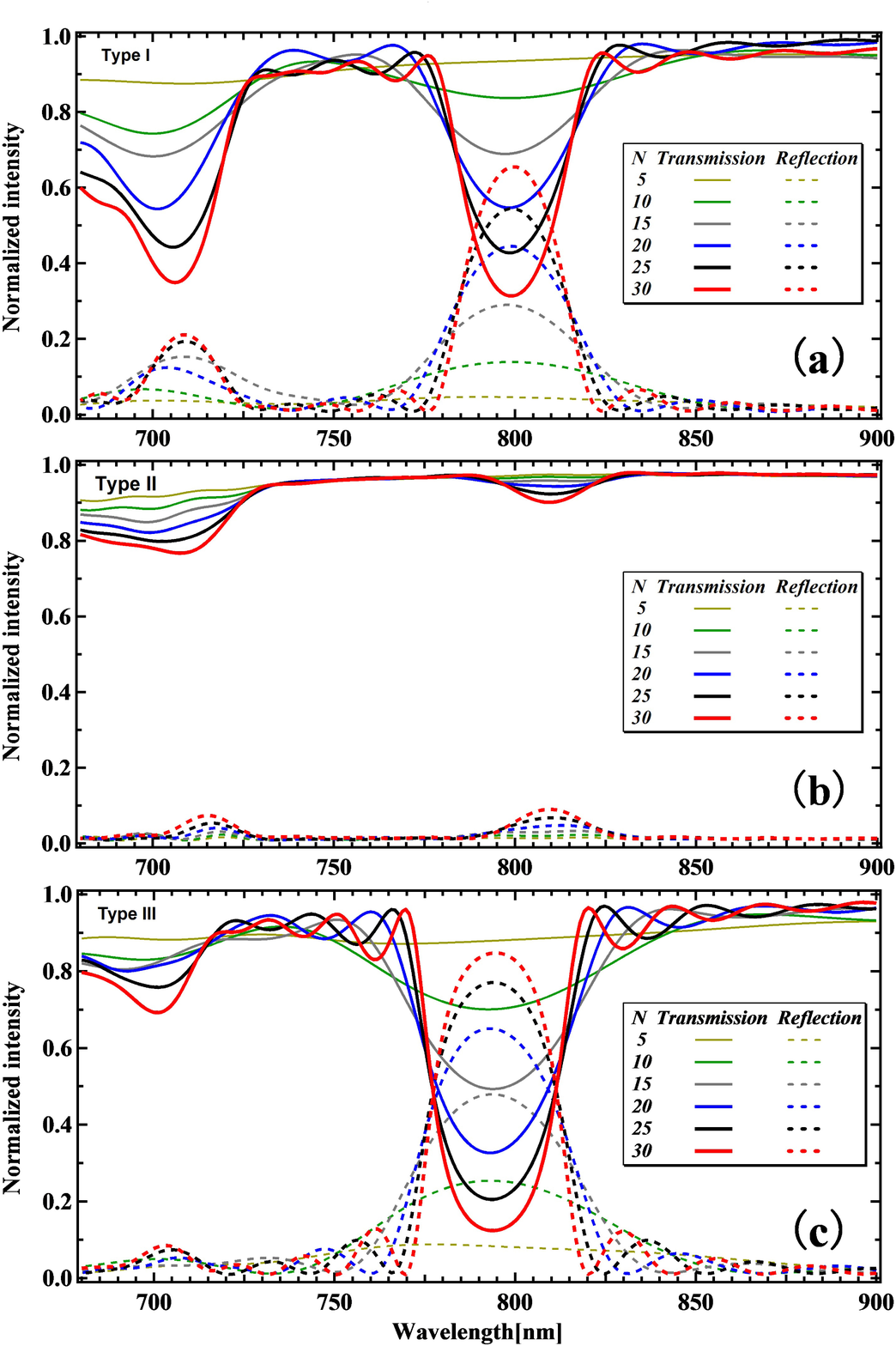}}
 \caption{Transmission and reflection spectra for different ONF structures for different period number (from 5 to 30). (a), (b) and (c) are the normalized transmission spectra for Type I, II and III structures, respectively. The solid lines correspond to the transmission spectra, which exhibit increasing transmission depth as the period number, $N$, increases. The dashed lines correspond to reflection spectra. }
\end{figure}

\begin{figure}[htb]
\centerline{
\includegraphics[width=8cm]{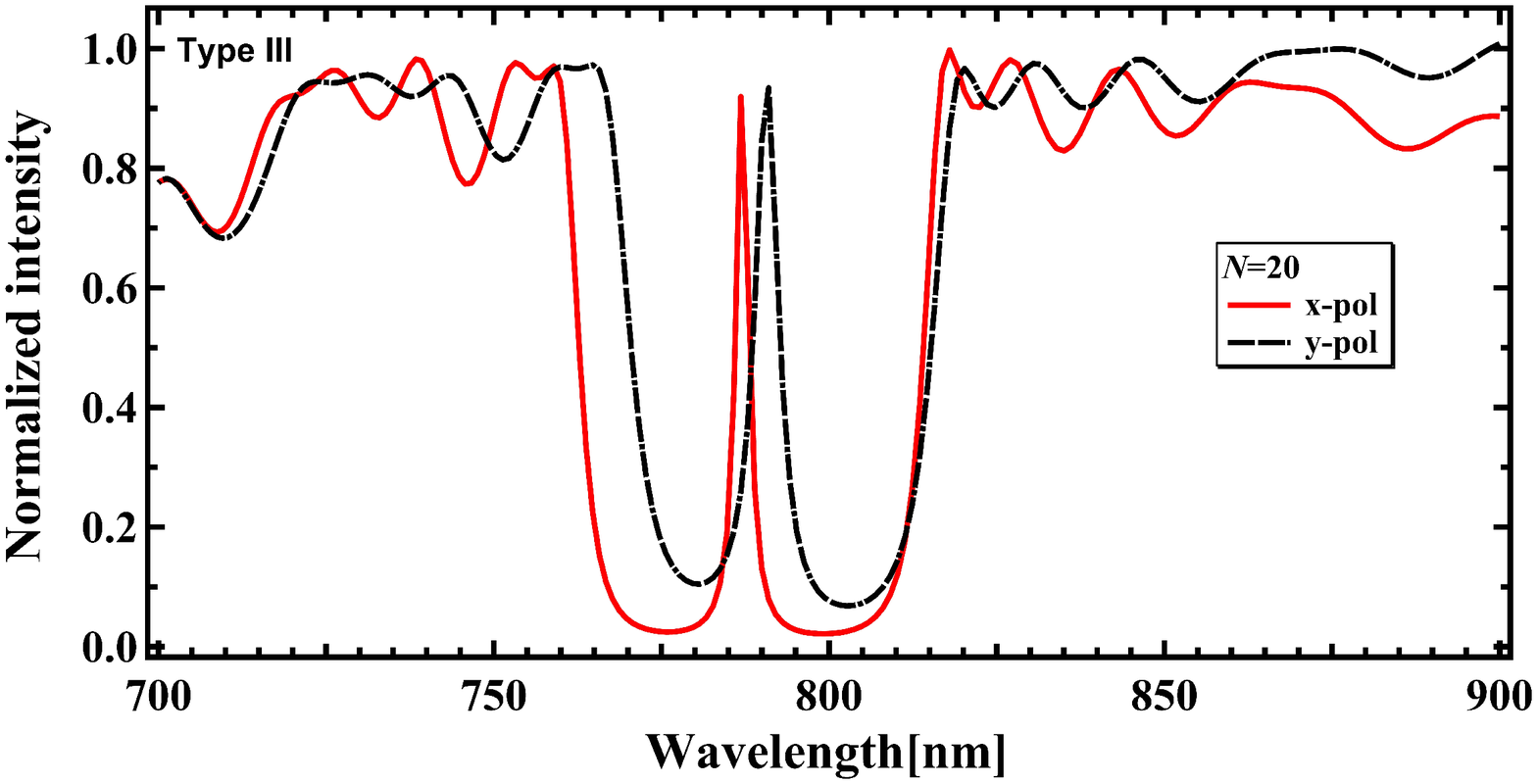}}
 \caption{Finite-different time-domain (FDTD)simulated transmission spectra of an ONF cavity with Type III structure. The red solid line represents $x$-polarization (x-pol) mode , whereas the dashed black line shows $y$-polarization (y-pol) mode with a periodic number $N = 20$ and a defect mirror distance of $ 2\ \upmu$m.}
\end{figure}

The simulation results indicate that some behaviors are common to each of the three structures. Transmission dips and reflection peaks become more prominent with increasing period number, $N$, as expected. The transmission and reflection band centered around 800 nm almost agree with initial design values.  Periodic fabrication in Type I and Type III lead to apparent changes in the optical spectra, whereas it has relatively little effect in the Type II structure. Moreover, the spectral width for the Type III structures is much wider and deeper than that for Type I and Type II. From these observations, the Type III cavity mirrors retain better optical characteristics with high reflectivity, caused by increased modulation of the etched-air units. In the following, we will focus on this type of cavity. 

Figure 3 shows the simulated transmission for a Type III cavity with $N = 20$, a cavity length of $ 2 \ \upmu$m  and an etched unit of $a \times b = 100 \times 100$ ${{\mathop{\rm nm}\nolimits} ^2}$, $c = 310$ nm and $D = 800$ nm. The existence of periodic air-filled etchings breaks the symmetry of the ONF at the waist, causing a strong polarization dependence of the electric field at this location. The resonance peak of the $x$-polarization (x-pol) mode  is narrower than that of the $y$-polarization (y-pol) mode as the light field due to the $x$-polarization experiences more modulation of the  refractive index.

\begin{figure}[htb]
\centerline{
\includegraphics[width=6cm]{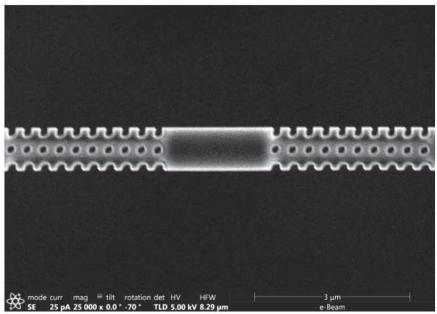}}
 \caption{SEM image of a cavity fabricated on an ONF with $N$ = 20. The fiber diameter is $\sim830\  $nm, milled-air square length is $\sim100$ nm, fabrication period is $\sim310$ nm and cavity length is $\sim2.2 \ \upmu$m.}	
\end{figure}
We fabricated the cavity using the following steps: ONFs are made by pulling a standard optical fiber (SM-800, FIBERCORE Inc.) over a hydrogen/oxygen flame until the central waist region is reduced to a diameter near or below that of the input laser wavelength.  A nanofiber with a diameter of $\sim$ 800 nm and a transmission of $\sim$ $94\% $ could be fabricated using the standard flame-brushing technique \cite{30}. The tapered fiber is then transferred to an FIB machine (Helios NanoLab DualBeam 650 system) in order to etch the Type III triplex periodic air cubes at the waist region. Before etching with the FIB, the ONF was coated with a conductive, transparent material i.e. indium titanium oxide (ITO) in order to mitigate charging effects. A 30 keV, 7 pA beam of ${\rm{G}}{{\rm{a}}^{\rm{ + }}}$ with a beam diameter of $\sim9$ nm was focused on the ITO-coated ONF for milling the structure. Scan electron microscopy (SEM) was used to measure the ONF diameters and to monitor the fiber during the milling process. A sample SEM image of an ONF cavity is given in Fig. 4, for a fiber diameter of $\sim830$ nm, a milled-air square length of $\sim100$ nm, a fabrication period of $\sim310$ nm and cavity length of $\sim2.2\ \upmu$m.
\begin{figure}[htb]
\centerline{
\includegraphics[width=8cm]{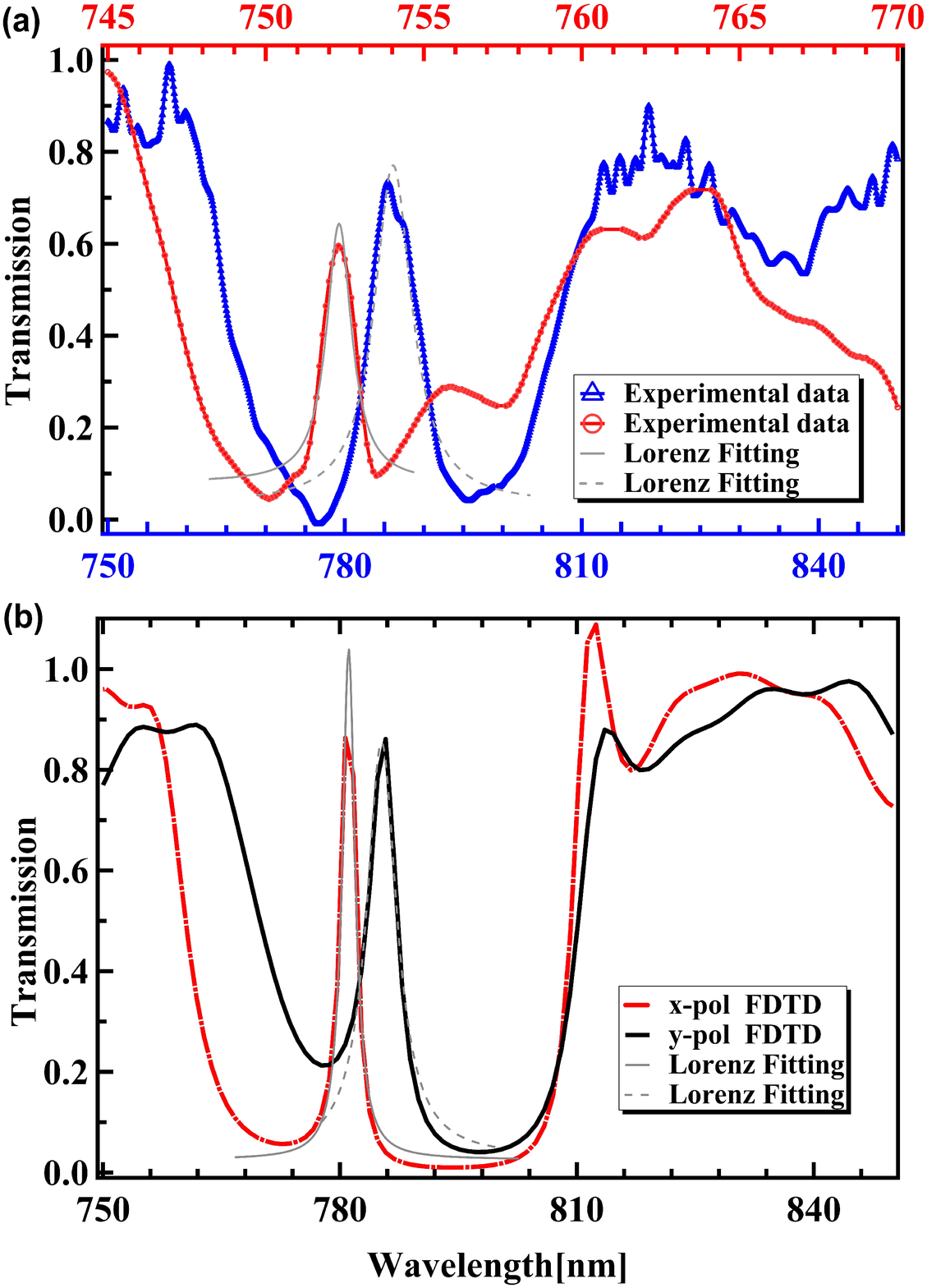}}
 \caption{(a) FDTD-simulated transmission spectra of an optical nanofiber cavity with Type III periodic structures. The red solid line corresponds to the $x$-polarization (x-pol) mode and the black line represents the $y$-polarization (y-pol) mode with a periodic number of $N=20$ and defect length of $ 2.2\ \upmu$m. (b) Corresponding experimental data. The spectra for the two orthogonally-polarized input modes are given by the red and blue lines. A Lorentz function is used to fit the full-width-at-half-maximum (FWHM), represented by the gray solid or dashed line.}
\end{figure}

After fabrication, the cavity was optically characterized by recording transmission spectra in the usual manner. A  super-continuum laser (NKT Photonics Inc.) was coupled to the fiber pigtails, and the cavity transmission from the fiber output was recorded on a laser spectrum analyzer (LSA) (HighFinesse Inc.). This was repeated for different input polarizations. By rotating a half wavelength plate, the input polarization was aligned with the two orthogonally-polarized cavity modes. Figure 5(a) shows the normalized cavity transmission spectrum for the two orthogonal polarizations (x-pol and y-pol mode). The corresponding FDTD simulation results are given in Fig. 5(b), where the red, solid line corresponds to the x-pol mode ($x$-direction) input and the black, solid line represents the y-pol mode ($y$-direction) input. The two sets of transmission spectra differ slightly. The resonance wavelengths measured in the experiment agree well with the rubidium ${D_2}$ line resonance simulation results. The full-width-at-half-maximum (FWHM) of the peaks can be determined using a Lorentz function fit, and for the FDTD simulations are found to be $1.96 \pm 0.12$ nm for a center wavelength of $781.1\pm 0.02$ nm and $4.64\pm 0.42$ nm for a wavelength of $785.27 \pm 0.10$ nm, while the results for the experiment are $0.96 \pm 0.10$ nm at the central wavelength of $752.32\pm 0.02$ nm and $5.65\pm 0.21$ nm at $786.11\pm 0.05$ nm. Based on these values, we estimate that the $Q$-factor of our cavity can reach ${\rm{784 \pm  87}}$. The simulation also indicates that the width of the transmission peak could be decreased to GHz or even MHz linewidth if the cavity length were increased. This compact design nanofiber cavity with mode volume of $\sim$1.05$\ \upmu$m$^3$ using the parameters in figure 4 shows great potential for application in strongly coupled solid-state, or even atomic, systems as a means of enhancing the interaction strength between single quantum emitters and single photons.

In summary, we have developed a compact nanofiber optical cavity using a focused ion beam technique. Mirror structures on an ONF exhibit a higher reflectivity than those of Bragg gratings and 1-D photonic crystal structures. The cavity's optical characteristics were investigated both experimentally and by FDTD simulation. The measured resonance wavelength is in good agreement with the theoretical resonance around the ${D_2}$ line. With a relatively short period, we estimate that the quality factor of the cavity structure could reach ${\rm{784 \pm  87}}$, which can be further improved by increasing the cavity length or mirror periods. These structures are show potential for studying cavity QED effects and for constructing large-scale quantum networks in the future.

This work was supported by Okinawa Institute of Science and Technology Graduate University. The authors would like to thank F. Le Kien for invaluable discussions. R. Murphy and S. Aird are acknowledged for comments on the manuscript.

\end{document}